\documentclass[amsmath,amssymb,aps,prd,groupedaddress,superscriptaddress,preprint]{revtex4-1}

\usepackage[thickspace,amssymb]{SIunits}
\usepackage{graphicx}
\usepackage{dcolumn}
\usepackage{bm}
\usepackage{lineno}
\usepackage{textcomp}

\begin{document}

\preprint{Preprint submitted to Phys. Rev. D}

\title{The cosmic ray proton plus helium energy spectrum measured by the ARGO--YBJ experiment in the energy range 3--300 TeV}

\author{B.~Bartoli}
\affiliation{Dipartimento di Fisica dell'Universit\`a di Napoli
                  ``Federico II'', Complesso Universitario di Monte
                  Sant'Angelo, via Cinthia, 80126 Napoli, Italy.}
\affiliation{Istituto Nazionale di Fisica Nucleare, Sezione di
                  Napoli, Complesso Universitario di Monte
                  Sant'Angelo, via Cinthia, 80126 Napoli, Italy.}

\author{P.~Bernardini}
\affiliation{Dipartimento Matematica e Fisica "Ennio De Giorgi", 
                  Universit\`a del Salento,
                  via per Arnesano, 73100 Lecce, Italy.}
\affiliation{Istituto Nazionale di Fisica Nucleare, Sezione di
                  Lecce, via per Arnesano, 73100 Lecce, Italy.}
\author{X.J.~Bi}
\affiliation{Key Laboratory of Particle Astrophysics, Institute
                  of High Energy Physics, Chinese Academy of Sciences,
                  P.O. Box 918, 100049 Beijing, P.R. China.}
\author{Z.~Cao}
\affiliation{Key Laboratory of Particle Astrophysics, Institute
                  of High Energy Physics, Chinese Academy of Sciences,
                  P.O. Box 918, 100049 Beijing, P.R. China.}
                  
\author{S.~Catalanotti}
 \affiliation{Dipartimento di Fisica dell'Universit\`a di Napoli
                  ``Federico II'', Complesso Universitario di Monte
                  Sant'Angelo, via Cinthia, 80126 Napoli, Italy.}
\affiliation{Istituto Nazionale di Fisica Nucleare, Sezione di
                  Napoli, Complesso Universitario di Monte
                  Sant'Angelo, via Cinthia, 80126 Napoli, Italy.}

\author{S.Z.~Chen}
\affiliation{Key Laboratory of Particle Astrophysics, Institute
                  of High Energy Physics, Chinese Academy of Sciences,
                  P.O. Box 918, 100049 Beijing, P.R. China.}
 
 \author{T.L.~Chen}
\affiliation{Tibet University, 850000 Lhasa, Xizang, P.R. China.}

 \author{S.W.~Cui}
\affiliation{Hebei Normal University, Shijiazhuang 050016,
                   Hebei, P.R. China.}

 \author{B.Z.~Dai}
\affiliation{Yunnan University, 2 North Cuihu Rd., 650091 Kunming,
                   Yunnan, P.R. China.}
                   
 \author{A.~D'Amone}
 \affiliation{Dipartimento Matematica e Fisica "Ennio De Giorgi", 
                  Universit\`a del Salento,
                  via per Arnesano, 73100 Lecce, Italy.}
\affiliation{Istituto Nazionale di Fisica Nucleare, Sezione di
                  Lecce, via per Arnesano, 73100 Lecce, Italy.}

\author{ Danzengluobu}
\affiliation{Tibet University, 850000 Lhasa, Xizang, P.R. China.}
 
\author{I.~De Mitri}
\affiliation{Dipartimento Matematica e Fisica "Ennio De Giorgi", 
                  Universit\`a del Salento,
                  via per Arnesano, 73100 Lecce, Italy.}
\affiliation{Istituto Nazionale di Fisica Nucleare, Sezione di
                  Lecce, via per Arnesano, 73100 Lecce, Italy.}

 \author{B.~D'Ettorre Piazzoli}
 \affiliation{Dipartimento di Fisica dell'Universit\`a di Napoli
                  ``Federico II'', Complesso Universitario di Monte
                  Sant'Angelo, via Cinthia, 80126 Napoli, Italy.}
\affiliation{Istituto Nazionale di Fisica Nucleare, Sezione di
                  Napoli, Complesso Universitario di Monte
                  Sant'Angelo, via Cinthia, 80126 Napoli, Italy.}

 \author{T.~Di Girolamo}
 \affiliation{Dipartimento di Fisica dell'Universit\`a di Napoli
                  ``Federico II'', Complesso Universitario di Monte
                  Sant'Angelo, via Cinthia, 80126 Napoli, Italy.}
\affiliation{Istituto Nazionale di Fisica Nucleare, Sezione di
                  Napoli, Complesso Universitario di Monte
                  Sant'Angelo, via Cinthia, 80126 Napoli, Italy.}

 \author{G.~Di Sciascio}
\affiliation{Istituto Nazionale di Fisica Nucleare, Sezione di
                  Roma Tor Vergata, via della Ricerca Scientifica 1,
                  00133 Roma, Italy.}

 \author{C.F.~Feng}
\affiliation{Shandong University, 250100 Jinan, Shandong, P.R. China.}

\author{ Zhaoyang Feng}
 \affiliation{Key Laboratory of Particle Astrophysics, Institute
                  of High Energy Physics, Chinese Academy of Sciences,
                  P.O. Box 918, 100049 Beijing, P.R. China.}

 \author{Zhenyong Feng}
 \affiliation{Southwest Jiaotong University, 610031 Chengdu,
                   Sichuan, P.R. China.}
                   
 \author{Q.B.~Gou}
 \affiliation{Key Laboratory of Particle Astrophysics, Institute
                  of High Energy Physics, Chinese Academy of Sciences,
                  P.O. Box 918, 100049 Beijing, P.R. China.}

 \author{Y.Q.~Guo}
 \affiliation{Key Laboratory of Particle Astrophysics, Institute
                  of High Energy Physics, Chinese Academy of Sciences,
                  P.O. Box 918, 100049 Beijing, P.R. China.}

 \author{H.H.~He}
 \affiliation{Key Laboratory of Particle Astrophysics, Institute
                  of High Energy Physics, Chinese Academy of Sciences,
                  P.O. Box 918, 100049 Beijing, P.R. China.}

 \author{Haibing Hu} 
\affiliation{Tibet University, 850000 Lhasa, Xizang, P.R. China.}
 
 \author{Hongbo Hu}
 \affiliation{Key Laboratory of Particle Astrophysics, Institute
                  of High Energy Physics, Chinese Academy of Sciences,
                  P.O. Box 918, 100049 Beijing, P.R. China.}

\author{M.~Iacovacci}
\affiliation{Dipartimento di Fisica dell'Universit\`a di Napoli
                  ``Federico II'', Complesso Universitario di Monte
                  Sant'Angelo, via Cinthia, 80126 Napoli, Italy.}
\affiliation{Istituto Nazionale di Fisica Nucleare, Sezione di
                  Napoli, Complesso Universitario di Monte
                  Sant'Angelo, via Cinthia, 80126 Napoli, Italy.}

 \author{R.~Iuppa}
\affiliation{Istituto Nazionale di Fisica Nucleare, Sezione di
                  Roma Tor Vergata, via della Ricerca Scientifica 1,
                  00133 Roma, Italy.}
\affiliation{Dipartimento di Fisica dell'Universit\`a di Roma 
                  ``Tor Vergata'', via della Ricerca Scientifica 1, 
                  00133 Roma, Italy.}

 \author{H.Y.~Jia}
\affiliation{Southwest Jiaotong University, 610031 Chengdu,
                   Sichuan, P.R. China.}
                   
\author{ Labaciren}
\affiliation{Tibet University, 850000 Lhasa, Xizang, P.R. China.}

 \author{H.J.~Li}
\affiliation{Tibet University, 850000 Lhasa, Xizang, P.R. China.}

 \author{C.~Liu}
 \affiliation{Key Laboratory of Particle Astrophysics, Institute
                  of High Energy Physics, Chinese Academy of Sciences,
                  P.O. Box 918, 100049 Beijing, P.R. China.}

 \author{J.~Liu}
\affiliation{Yunnan University, 2 North Cuihu Rd., 650091 Kunming,
                   Yunnan, P.R. China.}
                   
\author{ M.Y.~Liu}
 \affiliation{Tibet University, 850000 Lhasa, Xizang, P.R. China.}

\author{ H.~Lu}
 \affiliation{Key Laboratory of Particle Astrophysics, Institute
                  of High Energy Physics, Chinese Academy of Sciences,
                  P.O. Box 918, 100049 Beijing, P.R. China.}
                  
 \author{L.L.~Ma}
 \affiliation{Key Laboratory of Particle Astrophysics, Institute
                  of High Energy Physics, Chinese Academy of Sciences,
                  P.O. Box 918, 100049 Beijing, P.R. China.}
                  
 \author{X.H.~Ma}
 \affiliation{Key Laboratory of Particle Astrophysics, Institute
                  of High Energy Physics, Chinese Academy of Sciences,
                  P.O. Box 918, 100049 Beijing, P.R. China.}
                  
 \author{G.~Mancarella}
   \affiliation{Dipartimento Matematica e Fisica "Ennio De Giorgi", 
                  Universit\`a del Salento,
                  via per Arnesano, 73100 Lecce, Italy.}
\affiliation{Istituto Nazionale di Fisica Nucleare, Sezione di
                  Lecce, via per Arnesano, 73100 Lecce, Italy.}
 
 \author{S.M.~Mari} \email[Corresponding author: ]{stefanomaria.mari@uniroma3.it}
\affiliation{Dipartimento di Matematica e Fisica dell'Universit\`a ``Roma Tre'',
                   via della Vasca Navale 84, 00146 Roma, Italy.} 
\affiliation{Istituto Nazionale di Fisica Nucleare, Sezione di
                  Roma Tre, via della Vasca Navale 84, 00146 Roma, Italy.}

 \author{G.~Marsella}
  \affiliation{Dipartimento Matematica e Fisica "Ennio De Giorgi", 
                  Universit\`a del Salento,
                  via per Arnesano, 73100 Lecce, Italy.}
\affiliation{Istituto Nazionale di Fisica Nucleare, Sezione di
                  Lecce, via per Arnesano, 73100 Lecce, Italy.}

\author{ S.~Mastroianni}
 \affiliation{Istituto Nazionale di Fisica Nucleare, Sezione di
                  Napoli, Complesso Universitario di Monte
                  Sant'Angelo, via Cinthia, 80126 Napoli, Italy.}

 \author{P.~Montini}\email[Corresponding author: ]{paolo.montini@roma3.infn.it}
\affiliation{Dipartimento di Matematica e Fisica dell'Universit\`a ``Roma Tre'',
                   via della Vasca Navale 84, 00146 Roma, Italy.} 
\affiliation{Istituto Nazionale di Fisica Nucleare, Sezione di
                  Roma Tre, via della Vasca Navale 84, 00146 Roma, Italy.}

 \author{C.C.~Ning}
\affiliation{Tibet University, 850000 Lhasa, Xizang, P.R. China.}

\author{ L.~Perrone}
   \affiliation{Dipartimento Matematica e Fisica "Ennio De Giorgi", 
                  Universit\`a del Salento,
                  via per Arnesano, 73100 Lecce, Italy.}
\affiliation{Istituto Nazionale di Fisica Nucleare, Sezione di
                  Lecce, via per Arnesano, 73100 Lecce, Italy.}

\author{ P.~Pistilli}
\affiliation{Dipartimento di Matematica e Fisica dell'Universit\`a ``Roma Tre'',
                   via della Vasca Navale 84, 00146 Roma, Italy.} 
\affiliation{Istituto Nazionale di Fisica Nucleare, Sezione di
                  Roma Tre, via della Vasca Navale 84, 00146 Roma, Italy.}
                  
\author{ P.~Salvini}
\affiliation{Istituto Nazionale di Fisica Nucleare, Sezione di Pavia,
                   via Bassi 6, 27100 Pavia, Italy.}

\author{ R.~Santonico}
\affiliation{Istituto Nazionale di Fisica Nucleare, Sezione di
                  Roma Tor Vergata, via della Ricerca Scientifica 1,
                  00133 Roma, Italy.}
\affiliation{Dipartimento di Fisica dell'Universit\`a di Roma 
                  ``Tor Vergata'', via della Ricerca Scientifica 1, 
                  00133 Roma, Italy.}

\author{G. Settanta}
\affiliation{Dipartimento di Matematica e Fisica dell'Universit\`a ``Roma Tre'',
                   via della Vasca Navale 84, 00146 Roma, Italy.} 
 
 \author{P.R.~Shen}
  \affiliation{Key Laboratory of Particle Astrophysics, Institute
                  of High Energy Physics, Chinese Academy of Sciences,
                  P.O. Box 918, 100049 Beijing, P.R. China.}
                  
 \author{X.D.~Sheng}
  \affiliation{Key Laboratory of Particle Astrophysics, Institute
                  of High Energy Physics, Chinese Academy of Sciences,
                  P.O. Box 918, 100049 Beijing, P.R. China.}
                  
 \author{F.~Shi}
  \affiliation{Key Laboratory of Particle Astrophysics, Institute
                  of High Energy Physics, Chinese Academy of Sciences,
                  P.O. Box 918, 100049 Beijing, P.R. China.}
                  
 \author{A.~Surdo}
 \affiliation{Istituto Nazionale di Fisica Nucleare, Sezione di
                  Lecce, via per Arnesano, 73100 Lecce, Italy.}

 \author{Y.H.~Tan}
  \affiliation{Key Laboratory of Particle Astrophysics, Institute
                  of High Energy Physics, Chinese Academy of Sciences,
                  P.O. Box 918, 100049 Beijing, P.R. China.}
                  
 \author{P.~Vallania}
\affiliation{Istituto Nazionale di Fisica Nucleare,
                   Sezione di Torino, via P. Giuria 1, 10125 Torino, Italy.}
\affiliation{Dipartimento di Fisica dell'Universit\`a di 
                   Torino, via P. Giuria 1, 10125 Torino, Italy.}

 \author{S.~Vernetto}
\affiliation{Istituto Nazionale di Fisica Nucleare,
                   Sezione di Torino, via P. Giuria 1, 10125 Torino, Italy.}
\affiliation{Dipartimento di Fisica dell'Universit\`a di 
                   Torino, via P. Giuria 1, 10125 Torino, Italy.}

 \author{C.~Vigorito}
\affiliation{Istituto Nazionale di Fisica Nucleare,
                   Sezione di Torino, via P. Giuria 1, 10125 Torino, Italy.}
\affiliation{Dipartimento di Fisica dell'Universit\`a di 
                   Torino, via P. Giuria 1, 10125 Torino, Italy.}

\author{ H.~Wang}
  \affiliation{Key Laboratory of Particle Astrophysics, Institute
                  of High Energy Physics, Chinese Academy of Sciences,
                  P.O. Box 918, 100049 Beijing, P.R. China.}
                  
\author{ C.Y.~Wu}
  \affiliation{Key Laboratory of Particle Astrophysics, Institute
                  of High Energy Physics, Chinese Academy of Sciences,
                  P.O. Box 918, 100049 Beijing, P.R. China.}
                  
\author{ H.R.~Wu}
  \affiliation{Key Laboratory of Particle Astrophysics, Institute
                  of High Energy Physics, Chinese Academy of Sciences,
                  P.O. Box 918, 100049 Beijing, P.R. China.}
                  
\author{ L.~Xue}
\affiliation{Shandong University, 250100 Jinan, Shandong, P.R. China.}

 \author{Q.Y.~Yang}
\affiliation{Yunnan University, 2 North Cuihu Rd., 650091 Kunming,
                   Yunnan, P.R. China.}

\author{ X.C.~Yang}
\affiliation{Yunnan University, 2 North Cuihu Rd., 650091 Kunming,
                   Yunnan, P.R. China.}

\author{ Z.G.~Yao}
  \affiliation{Key Laboratory of Particle Astrophysics, Institute
                  of High Energy Physics, Chinese Academy of Sciences,
                  P.O. Box 918, 100049 Beijing, P.R. China.}
                  
 \author{A.F.~Yuan}
\affiliation{Tibet University, 850000 Lhasa, Xizang, P.R. China.}

 \author{M.~Zha}
  \affiliation{Key Laboratory of Particle Astrophysics, Institute
                  of High Energy Physics, Chinese Academy of Sciences,
                  P.O. Box 918, 100049 Beijing, P.R. China.}
 
 \author{H.M.~Zhang}
  \affiliation{Key Laboratory of Particle Astrophysics, Institute
                  of High Energy Physics, Chinese Academy of Sciences,
                  P.O. Box 918, 100049 Beijing, P.R. China.}
 
 \author{L.~Zhang}
\affiliation{Yunnan University, 2 North Cuihu Rd., 650091 Kunming,
                   Yunnan, P.R. China.}
                   
 \author{X.Y.~Zhang}
\affiliation{Shandong University, 250100 Jinan, Shandong, P.R. China.}

 \author{Y.~Zhang}
  \affiliation{Key Laboratory of Particle Astrophysics, Institute
                  of High Energy Physics, Chinese Academy of Sciences,
                  P.O. Box 918, 100049 Beijing, P.R. China.}
 \author{J.~Zhao}
  \affiliation{Key Laboratory of Particle Astrophysics, Institute
                  of High Energy Physics, Chinese Academy of Sciences,
                  P.O. Box 918, 100049 Beijing, P.R. China.}

 \author{Zhaxiciren}
\affiliation{Tibet University, 850000 Lhasa, Xizang, P.R. China.}

 \author{Zhaxisangzhu}
\affiliation{Tibet University, 850000 Lhasa, Xizang, P.R. China.}

 \author{X.X.~Zhou}
\affiliation{Southwest Jiaotong University, 610031 Chengdu,
                   Sichuan, P.R. China.}

 \author{F.R.~Zhu}
\affiliation{Southwest Jiaotong University, 610031 Chengdu,
                   Sichuan, P.R. China.}

 \author{Q.Q.~Zhu}
  \affiliation{Key Laboratory of Particle Astrophysics, Institute
                  of High Energy Physics, Chinese Academy of Sciences,
                  P.O. Box 918, 100049 Beijing, P.R. China.}

\collaboration{ARGO--YBJ Collaboration}
\noaffiliation

\date{\today}
\begin{abstract}
The ARGO--YBJ experiment is a full--coverage air shower detector located at the Yangbajing Cosmic Ray Observatory (Tibet, People's Republic of China, 4300 m a.s.l.). The high altitude, combined with the full--coverage technique, allows the detection of extensive air showers in a wide energy range and offer the possibility of measuring the cosmic ray proton plus helium  spectrum down to the TeV region, where direct balloon/space--borne measurements are available.  The detector has been in stable data taking in its full configuration from November 2007 to February 2013. In this paper the measurement of the cosmic ray proton plus helium energy spectrum is presented in the region $3-300$ TeV by analyzing the full collected data sample. The resulting spectral index is $\gamma = -2.64 \pm 0.01$. These results demonstrate the possibility of performing an accurate measurement of the spectrum of light elements with a ground based air shower detector. 

\end{abstract}


\maketitle

\section{Introduction}
\label{intro}
Cosmic rays are ionized nuclei reaching the Earth from outside the solar system. Many experimental efforts have been devoted to the study of cosmic ray properties. In the last decades many experiments were focused on the identification of cosmic ray sources and on the understanding of their acceleration and propagation mechanisms. Despite a very large amount of data collected so far, the origin and propagation of cosmic rays are still under discussion. Supernova remnants (SNRs) are commonly identified as the source of galactic cosmic rays since they could provide the amount of energy needed in order to accelerate particles up to the highest energies in the Galaxy. The measurement of the diffuse gamma--ray radiation in the energy range $1-100 \, \giga \electronvolt$ supports these hypotheses on the origin and propagation of cosmic rays \cite{fermi_diffuse}.  Moreover the TeV gamma--ray emission from SNRs, detected by ground--based experiment, can be related to the acceleration of particles up to $\sim 100$ TeV \cite{hess2004, hess2007}. A very detailed measurement of the energy spectrum and composition of primary cosmic rays will lead to a deeper knowledge of  the acceleration and propagation mechanisms. Since the energy spectrum spans a huge energy interval, experiments dedicated to the study of cosmic ray properties are essentially divided into two broad classes. Direct experiments operating on satellites or balloons are able to measure the energy spectrum and the isotopic composition of cosmic rays on top of the atmosphere. Due to their reduced detector active surface and the limited exposure time the maximum detectable energy is limited up to few TeV. New generation instruments, capable of long balloon flights, have extended the energy measurements up to $\sim 100$ TeV. All the information concerning cosmic rays above 100 TeV is provided by ground--based air shower experiments. Air shower experiments are able to observe the cascade of particles produced by the interaction between cosmic rays and the Earth's atmosphere. Ground based experiments detect extensive air showers produced by primaries with energies up to $10^{20} \, \electronvolt$, however they do not allow an easy determination of the abundances of individual elements and the measurement of the composition is therefore limited only to the main elemental groups. Moreover, due to a lack of a model--independent energy calibration, the determination of the primary energy relies on the hadronic interaction model used in the description of the shower's development.\\
The ARGO--YBJ experiment is a high--altitude full--coverage air shower detector which was in full and stable data taking from November 2007 up to February 2013.  As described in section \ref{sec:argo}, the detector is equipped with a digital and an analog readout systems working independently in order to study the cosmic ray properties in the energy range $1-10^4 \,\tera \electronvolt$, which is one of the main physics goals of the ARGO--YBJ experiment. The high  space--time resolution of the digital readout system allows the detection of showers produced by primaries down to few TeV, where balloon--borne measurements are available. The analog readout system was designed and built in order to detect showers in a very wide range of particle density at ground level and to explore the cosmic ray spectrum up to the PeV region.
In 2012 a first measurement of the cosmic ray proton plus helium (light component) spectrum obtained by analyzing a small sample collected during the first period of data taking with the detector in its full configuration (by using the digital readout information only) has been presented \cite{argoPRD}. \\
In this paper we report the analysis of the full data sample collected by the ARGO--YBJ experiment in the period from January 2008 to December 2012  and the measurement of the light component energy spectrum of cosmic rays in the energy range $3-300\, \tera\electronvolt$ by applying an unfolding procedure based on the bayesian probabilities.  The analysis of the analog readout data and the corresponding cosmic ray spectrum up to the PeV energy region is in progress and will be addressed in a future paper.

\section{The ARGO-YBJ experiment}
\label{sec:argo}
The ARGO--YBJ experiment  (Yangbajing Cosmic Ray Observatory, Tibet, P.R. China. 4300 m a.s.l.) is a full--coverage detector made of a single layer of Resistive Plate Chambers (RPCs) with $\sim 93\%$ active area \cite{argo2006,argo2009}, surrounded by a partially instrumented guard ring designed to improve the event reconstruction. The detector is made of 1836 RPCs, arranged in 153 clusters each made of 12 chambers.  The digital readout consists of 18360 pads each segmented in 8 strips. A dedicated procedure was implemented to calibrate the detector in order to achieve high pointing accuracy  \cite{calib}. The angular and core reconstruction resolution are respectively  $0.4^\circ$  and 5 m for events with at least 500 fired pads \cite{moon,pinoicrc}. The installation of the central carpet was completed in June 2006. The guard ring was completed during spring 2007 and connected to the data acquisition system \cite{daq} in November 2007. A simple trigger logic based on the coincidence between the pad signals was implemented. The detector has been in stable data taking in its full configuration for more than five years with a trigger threshold $N_{pad} = 20$, corresponding to a trigger rate of about $3.6 \, \kilo\hertz$ and a dead time of $4\%$. The high granularity and time resolution of the detector provide a detailed three-dimensional reconstruction of the shower front. The high altitude location and the segmentation of the experiment offer the possibility to detect showers produced by charged cosmic rays with energies down to few TeV. The digital readout of the pad system allows reconstruction of showers with a particle density at ground level up to about 23 particles/$\meter^{2}$, which correspond to primaries up to a few hundreds of TeV. In order to extend the detector operating range and investigate energies up to the PeV region each RPC has been equipped with two large size electrodes called Big Pads \cite{analog-paper}. Each Big Pad provide a signal whose amplitude is proportional to the number of particles impinging the detector surface. The analog readout system allows a detailed measurement of showers with particle density at ground up to more than $10^{4}\, \mathrm{particles}/\meter^{2}$.

\section{Data analysis}
\label{sect:data_an}

\subsection{Unfolding of the cosmic ray energy spectrum}
\label{sect:bayes}
As widely described in \cite{thepaper,argoPRD}, the determination of the cosmic ray spectrum starting from the measured space--time distribution of charged particles at ground level is a classical unfolding problem that can be dealt by using the bayesian technique\cite{dago}.  In this framework the detector response is represented by the probability $P(M_j|E_i)$ of measuring a multiplicity $M_j$ due to a shower produced by a primary of energy $E_i$. The estimated number of events in a certain energy bin $E_i$ is therefore related to the number of events measured in a multiplicity bin $M_j$ by the equation
\begin{equation}
\label{eq:N_E}
\hat N({E_i}) \propto \sum_j N({M_j})P(E_i|M_j)
\end{equation}
where $\eta_{ij}$ is constructed by using the Bayes theorem
\begin{equation}
P(E_i|M_j) = \frac{P(M_j|E_i)P(E_i)}{\sum_k P(M_j|E_k)P(E_k)}.
\end{equation}
The values of the probability $P(M_j|E_i)$ are evaluated by means of a Monte Carlo simulation of the development of the shower and of the detector response. The quantity $P(E_i|M_j)$ represents the probability that a shower detected with multiplicity $M_j$ has been produced by a primary of energy $E_i$. The values of $P(M_j|E_i)$ are evaluated by means of an iterative procedure starting from a prior value $P^{(0)}(E_i)$,  in which in the $n$--th step $P^{(n)}(E_i)$ is replaced by the updated value
\begin{equation}
P^{(n+1)}(E_i) = \frac{\hat N^{(n)}(E_i)}{\sum_k \hat N^{(n)}(E_k)},
\end{equation} 
where $\hat N^{(n)}(E_i)$ is evaluated in the $n$-th step according to eq. \ref{eq:N_E}. As initial prior $P^{(0)}(E_i) \sim E^{-2.5}$ was chosen, the effect of using different prior distributions has been evaluated as negligible. The iterative procedure ends when the variation of all $\hat N(E_i)$ in two consecutive steps are evaluated as negligible, namely less than 0.1 \%. Typically the convergence is reached after 3 iterations.

\subsection{Air shower and detector simulations}
\label{sect:mc}
The development of the shower in the Earth's atmosphere has been simulated by using the CORSIKA (v. 6980) code \cite{corsika}. The electromagnetic component are described by the EGS4 code \cite{egs4,egs4b}, while the high energy hadronic interactions are reproduced by QGSJET-II.03 model \cite{qgsjet2006a, qgsjet2006b}. Low energy hadronic interactions are described by the FLUKA package \cite{fluka2006a, fluka2005}. Showers produced by Protons, Helium, CNO nuclei and Iron have been generated with a spectral index $\gamma = -1$ in the energy range $(0.316 - 3.16\times 10^4)$ TeV.  About $5\times 10^{7}$ showers have been generated in the zenith angle range 0--45 degrees and azimuth angle range 0-360 degrees. Showers were sampled at the Yangbajing altitude and the shower core was randomly distributed over an area of $(250 \times 250)\, \meter^{2}$ centered on the detector.  The resulting CORSIKA showers have been processed by a GEANT3 \cite{geant3} based code in order to reproduce the detector response, including the effects of time resolution, RPC efficiency, trigger logic, accidental background produced by each pad and electronic noise.

\subsection{Event selection}
\label{sect:sel}

The ARGO-YBJ experiment was in stable data taking in its full configuration for more than five years: more than $5 \times 10^{11}$ events have been recorded and reconstructed. Several tools have been implemented in order to monitor the detector operation and reconstruction quality. The detector control system (DCS) \cite{camarri2003} continuously monitors the RPC current, the high voltage distribution, the gas mixture and the environmental conditions (temperature, pressure, humidity). In this work  the analysis of events collected during the period 2008--2012 is presented. Data and simulated events have been selected according to a multi--step procedure in order to obtain high quality events and to ensure a reliable and unbiased evaluation of the bayesian probabilities. The first step concerns the run  selection: in order to obtain a sample of high--quality runs,  the working condition of the detector and the quality of the reconstruction procedure have been analyzed by using the criteria described below.

\begin{figure}[h]
\begin{center}
\includegraphics[width =0.42\textwidth]{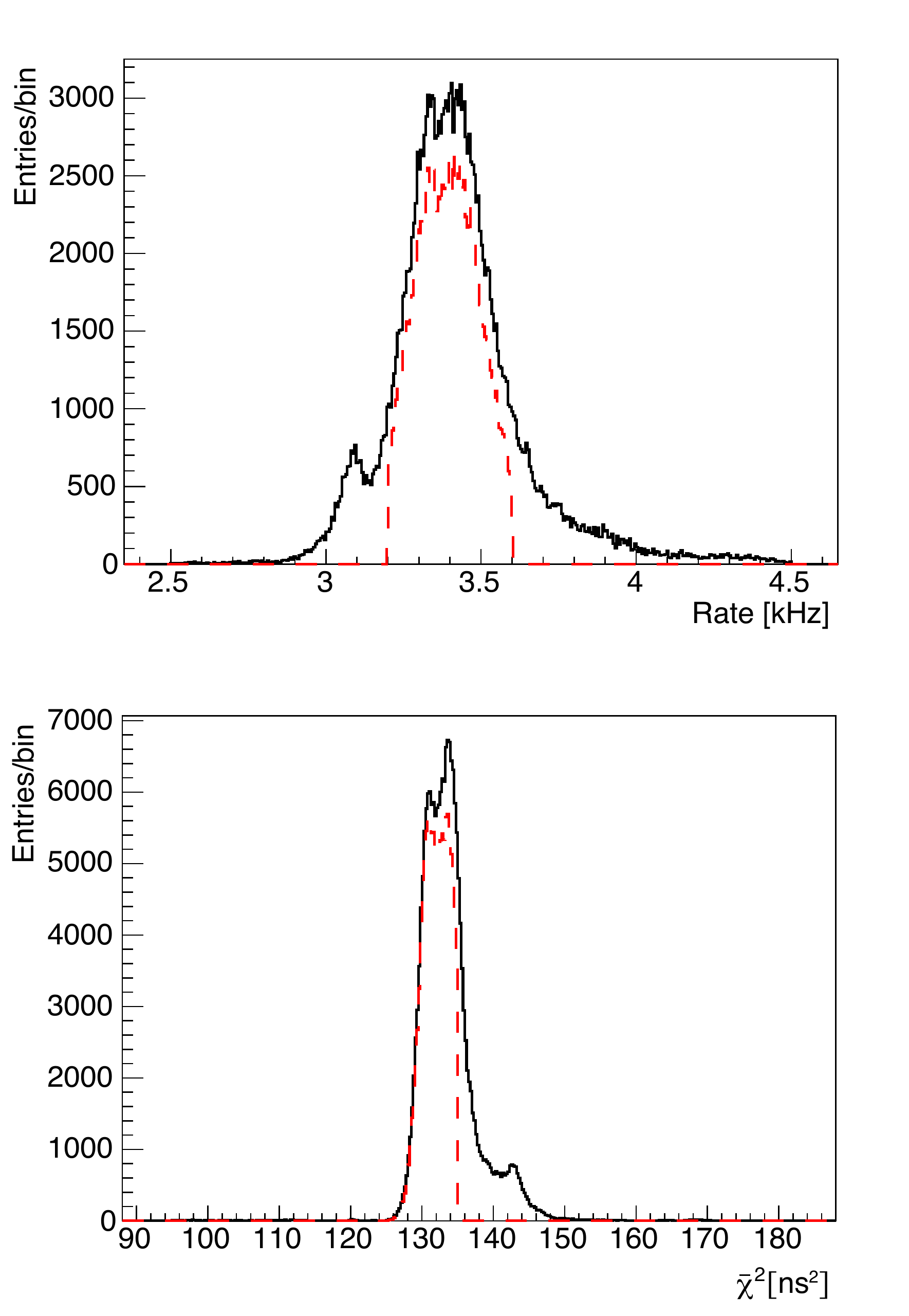}
\caption{Distribution of the trigger rate (top)  and of the unnormalised $\bar \chi^2$ (bottom) of all the runs collected by the ARGO--YBJ experiment (black lines). The resulting 2008-2012 sample selected according to the criteria described in section \ref{sect:data_an} is also reported (dashed red lines)}
\label{fig:trig-rate}
\end{center}
\end{figure}

\begin{itemize}
\item At least 128 clusters out of 130 must be active and connected to the DAQ and trigger systems. This criterium selects runs taken with almost the whole apparatus in data taking, discarding the runs that can bias the analysis because of the switched--off clusters.
\item Only runs with a duration $T \geqslant 1800 \, \second$ have been considered. The runs with a short duration are generally produced when a problem in the apparatus occurs. These runs have been removed from the analysis.
\item The value of the trigger rate for each run must stay within the range $3.2 - 3.7\, \kilo \hertz$. A trigger rate outside this range usually indicates that the detector was not operated standardly. These runs have been discarded.
\item To monitor the quality of the event reconstruction the mean value of the unnormalised $\chi^2$  obtained by fitting the shower front must be less than $135 \, \nano \second^2$ (see figure \ref{fig:trig-rate}). Nearly all runs that have $\bar \chi^2 > 135 \, \nano \second^2$ encountered some sort of problems.
\end{itemize}
In figure \ref{fig:trig-rate} the distribution of the trigger rate and the $\bar \chi^2$ of the reconstruction procedure are reported.  The procedure described above selects a data sample of about $3 \times 10^{11}$ events, corresponding to a live time of about 24000 hours.

\begin{figure}[t]
\includegraphics[width =0.41\textwidth]{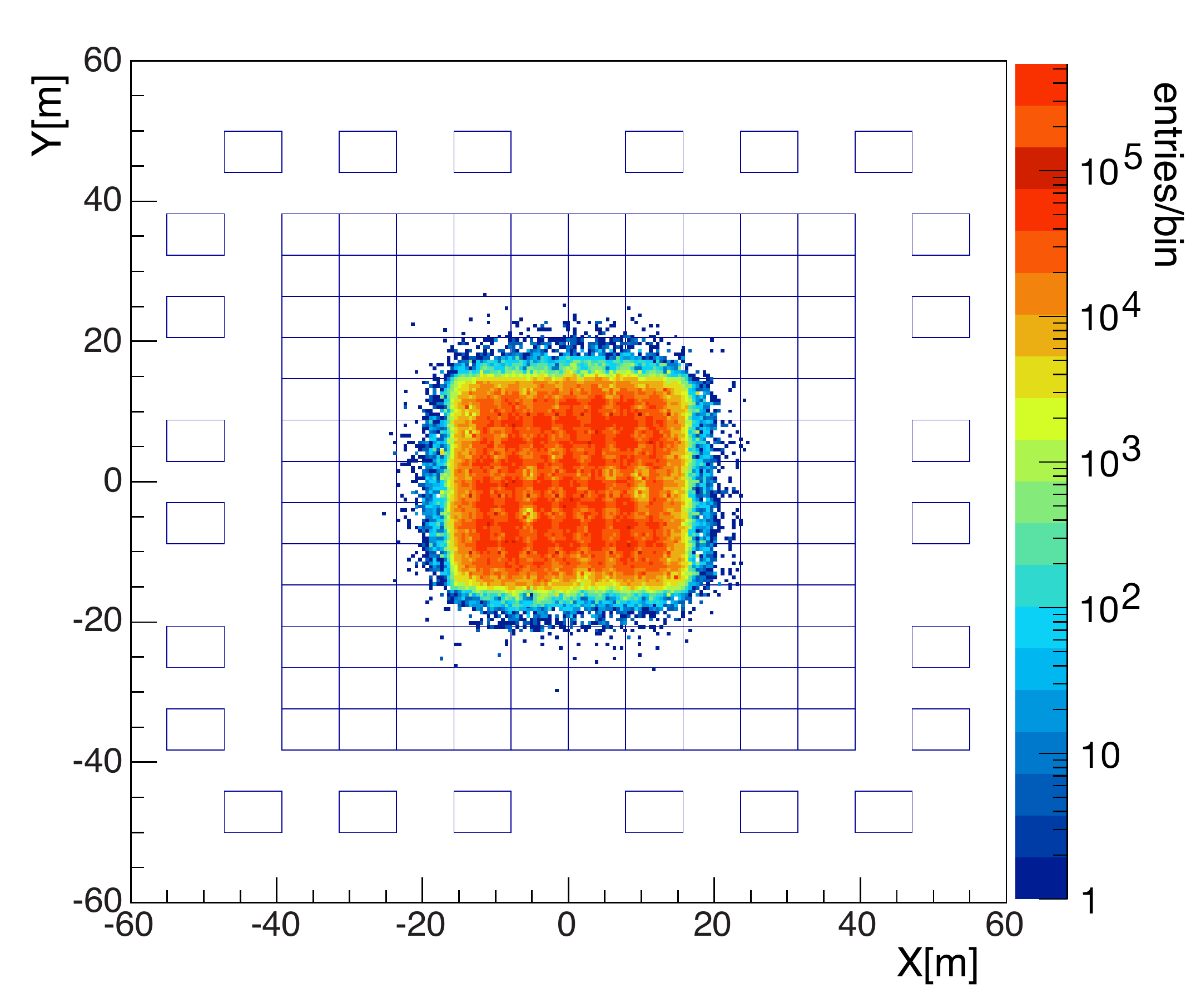}
\caption{Distribution of reconstructed core positions of showers selected by applying the criteria described in section \ref{sect:sel}.  The boxes represent the clusters layout. \label{fig:core}}
\end{figure}

\noindent
The following selection criteria (fiducial cuts) have been applied to both Monte Carlo and experimental data in order to improve the quality of the reconstruction and to obtain the best estimation of the bayesian probabilities.

\begin{itemize}
\item Only events with reconstructed zenith angles $\vartheta_R \leqslant 35^\circ$ have been considered. The resulting solid angle $\Omega$ is about 1.13 sr.
\item The measured shower multiplicity $M$ had to be in the range $150 \leqslant M \leqslant 5\times 10^4$. This selection cut was introduced in order to reduce bias effects in the estimation of the bayesian probabilities that are mainly located at the edges of the simulated energy range. Moreover the highest multiplicity cut avoid saturation effects of the digital readout system.
\item The cluster with the highest multiplicity had to be contained within an area of about $40 \times 40\, \meter^2$ centered on the detector. This cut was applied in order to reject events with their true shower core position located outside the detector surface.
\end{itemize}
In order to select showers induced by proton and helium nuclei the following criterium has been used.
\begin{itemize}
\item \emph{Density cut:} the average particle density ($\rho_{in}$) measured by the central area (20 inner clusters) of the detector must be higher than the particle density ($\rho_{out}$) measured by the outermost area (42 outer clusters): ($\rho_{in} > 1.25\,  \rho_{out}$). This selection criteria based on the lateral particle distribution was introduced in order to discard events produced by nuclei heavier than helium. In fact, in showers induced by heavy primaries the lateral distribution is wider than in light--induced ones.  By applying this criterion on events with the core located in a narrow area around the detector center, showers mainly produced by light primaries have been selected. The contamination of elements heavier than helium does not exceed few \%, as discussed in section \ref{sect:cont}.
\end{itemize}

\begin{figure}[t]
\includegraphics[width =0.4\textwidth]{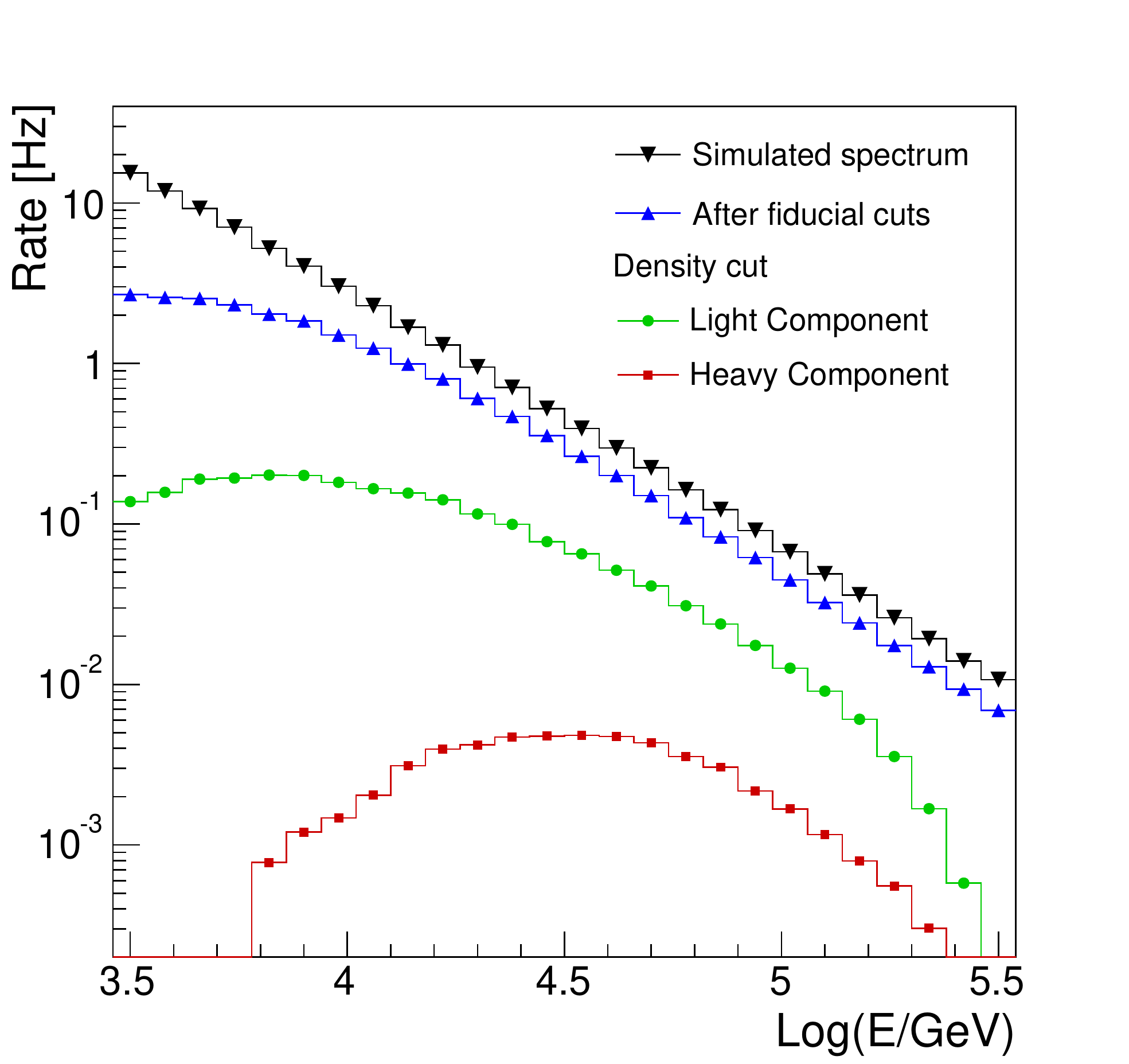}
\caption{Energy distribution of all Monte Carlo events (black) and of those surviving the fiducial cuts (blue) and the density cut (green and red) described in section \ref{sect:sel} according to the H\"orandel model \cite{horandel2003}. 
\label{fig:eff}}
\end{figure}

\begin{figure*}[t]
\includegraphics[width =0.81\textwidth]{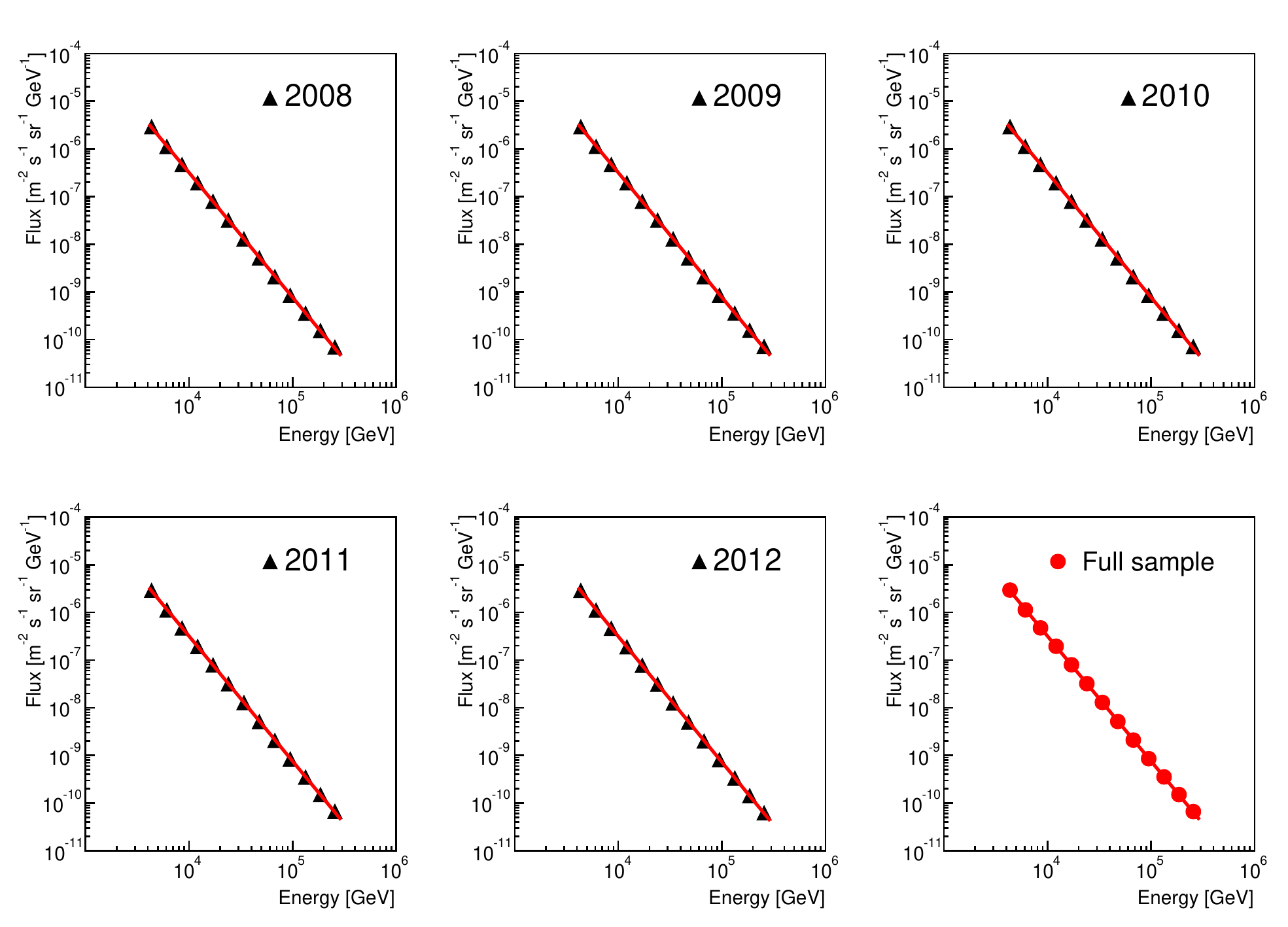} 
\caption{The light component spectrum measured by the ARGO--YBJ experiment by using data taken in each year of the period 2008--2012 and the full 2008--2012 data sample. The power--law fit of each spectrum is also reported (red lines).  \label{fig:spec_allyear}}
\end{figure*}

In figure \ref{fig:core} the coordinates of the reconstructed core position of the events surviving the selection criteria described above are reported. The plot shows that the contribution of events located outside an area of $40 \times 40 \, \meter^2$ is negligible. In figure \ref{fig:eff} the event rate obtained by using the H\"orandel model for input spectra and isotopic composition \cite{horandel2003} and surviving the selection criteria described above is reported as a function of energy for both proton plus helium (light component) and heavier elements (heavy component). The plot shows that the selected sample is essentially made of light nuclei. 

\section{The light component spectrum}
\label{sect:spectrum}
\begin{figure}[t]
\includegraphics[width =0.32\textwidth]{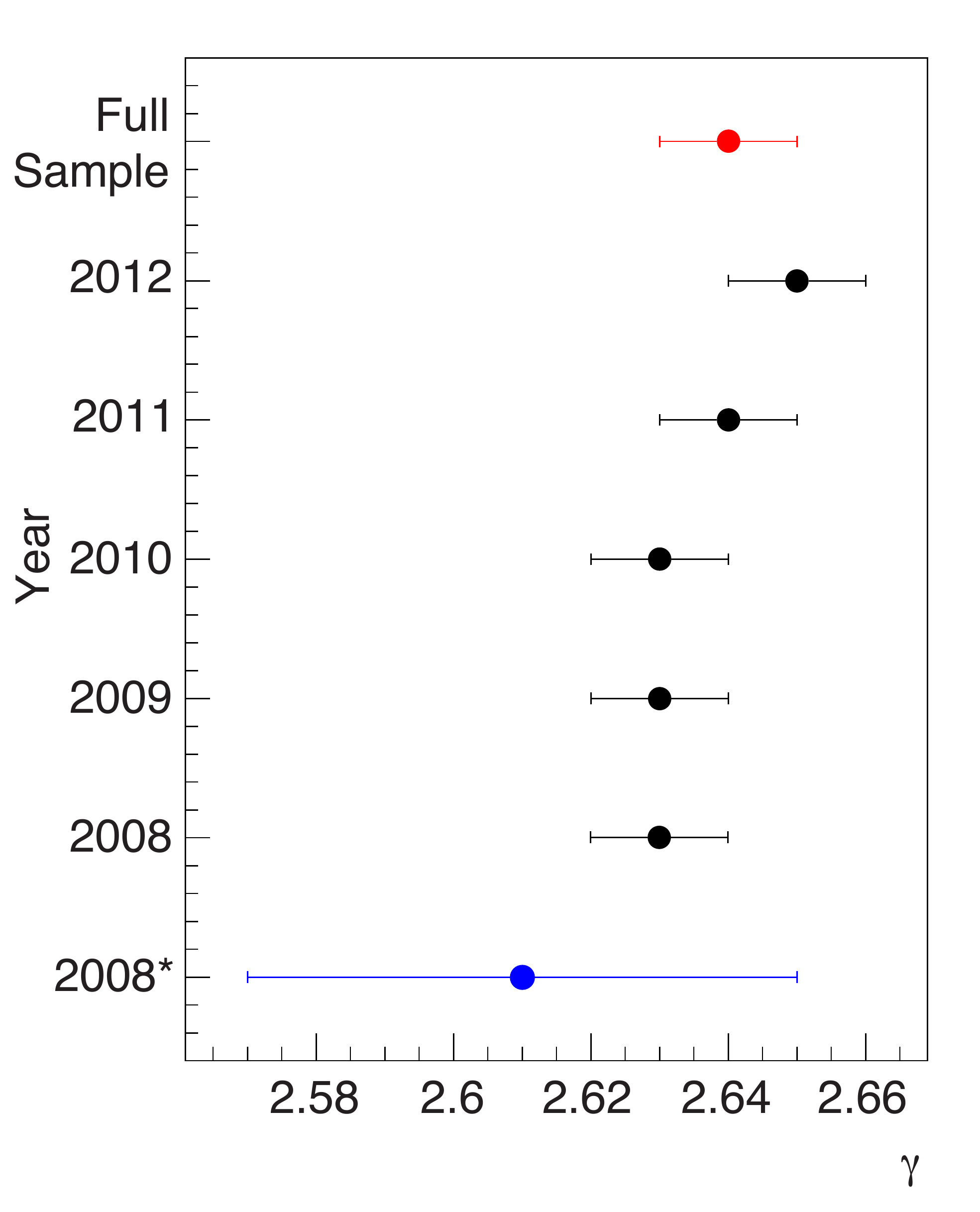}
\caption{Spectral indices of the power--law fit of the light component spectrum measured by analyzing the data sample collected in the period 2008--2012. The spectral index obtained in a previous analysis of the ARGO--YBJ data is shown as 2008* \cite{argoPRD} \label{fig:gamma}. The error bars represent the total uncertainty.}
\end{figure}

\begin{figure*}
\includegraphics[width =0.77\textwidth]{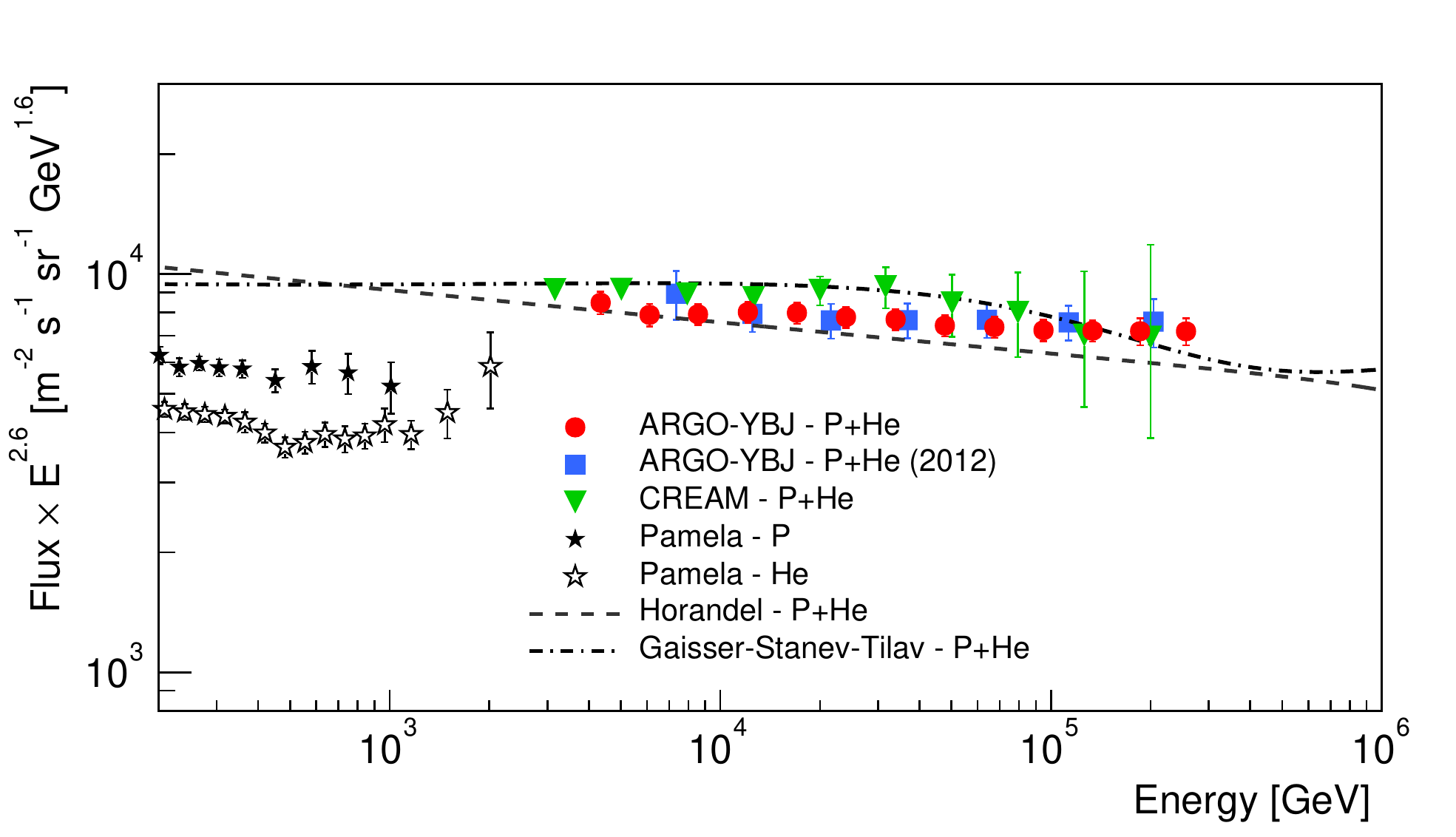}
\caption{The proton plus helium spectrum measured by the ARGO--YBJ experiment using the full 2008--2012 data sample. The error bars represent the total uncertainty.  Previous measurement performed by ARGO--YBJ in a narrower energy range by analyzing a smaller data sample is also reported (blue squares) \cite{argoPRD}. The green inverted triangles represent the sum of the proton and helium spectra measured by the CREAM experiment \cite{cream-apj2011}. The proton (stars) and helium (empty stars) spectra measured by the PAMELA experiment \cite{pamela} are also shown. The  light component spectra according to the Gaisser-Stanev-Tilav (dashed--dotted line) \cite{GST} and H\"orandel (dashed line) \cite{horandel2003} models are also shown. \label{fig:spectrum-full}}
\end{figure*}

The analysis was performed on the sample selected by the criteria described in section \ref{sect:data_an}. Simulated events have been sorted in 16 multiplicity bins and 13 energy bins in order to minimize the statistical error and to reduce bin migration effects. The Monte Carlo data sample was analyzed in order to evaluate the probability distribution $P(M|E)$ and the energy resolution which turns out to be about 10\%  for energies below 10 $\tera \electronvolt$ and of the order of 5\% at energies of about 100 $\tera \electronvolt$. The multiplicity distribution extracted from data has been unfolded according to the procedure described in section \ref{sect:bayes}. Results are reported in figure \ref{fig:spec_allyear} for each year of data taking and also for the full sample. In order to investigate the stability of the detector over a long period the analysis was performed separately on the data samples collected during each solar year in the period 2008--2012. The values of the proton plus helium flux measured at $50$ TeV are reported in table \ref{tbl:flux50tev}. A power--law fit has been performed on the measured spectrum of each year and of the full data sample, the resulting spectral indices are reported in figure \ref{fig:gamma}. Both the spectral indices and the flux values are in very good agreement between them, demonstrating the long--period reliability and the stability of the detector. The spectral index $\gamma = -2.64 \pm 0.01$, obtained by analyzing the full data sample, is in good agreement with the one measured by using a smaller data sample collected in the first months of 2008 \cite{argoPRD} which was not corrected by the contamination from heavier nuclei (see section \ref{sect:cont}).  \\
In table \ref{tab:flux} and figure \ref{fig:spectrum-full} the flux obtained by analyzing the full data sample is reported. The spectrum  covers a wide energy range, spanning about two orders of magnitude and is in excellent agreement with the previous ARGO--YBJ measurement.  Statistical errors are of the order of 1\textperthousand, more than $10^5$ events have been selected in the highest energy region, while at the lowest energies more than $10^7$ events have been selected. Systematic errors are discussed in the next section.  The ARGO--YBJ data are in good agreement with the CREAM proton plus helium spectrum \cite{cream-apj2011}. At energies around 10 TeV and 50 TeV the fluxes differ by about 10\% and 20\% respectively. This means that the absolute energy scale difference of the two experiments is within 4\% and 6\%. The uncertainty on the absolute energy scale has been evaluated by exploiting the Moon shadow tool at a level of 10\% for energies below 30 TeV \cite{moon}. At present the ARGO--YBJ experiment is the only ground--based detector able to investigate the cosmic ray energy spectrum in this energy region. 

\begin{table}[h] 
\caption{\label{tbl:flux50tev} Proton plus helium flux measured at $5.0 \times 10^4$ GeV. }.
\begin{ruledtabular}
\begin{tabular}{cc}
Year & Flux $\pm$ tot. error $[ \meter^{-2} \second^{-1} \mathrm{sr}^{-1} \giga \electronvolt^{-1}]$\\
2008 & $(4.53\pm 0.28) \times 10^{-9}$ \\
2009 & $(4.54\pm 0.28) \times 10^{-9}$ \\
2010 & $(4.54\pm 0.28) \times 10^{-9}$ \\
2011 & $(4.50\pm 0.27) \times 10^{-9}$ \\
2012 & $(4.36 \pm 0.27) \times 10^{-9}$ \\
\end{tabular}
\end{ruledtabular}
\end{table}

\begin{table}[] 
\caption{\label{tab:flux} Light component energy spectrum measured by the ARGO--YBJ experiment by using the full 2008--2012 data sample in each energy bin.}
\begin{ruledtabular}
\begin{tabular}{c|c|c}
Energy Range & Energy & Flux $\pm$ total error\\
$[\giga\electronvolt]$ & $[\giga\electronvolt]$ & $[ \meter^{-2} \second^{-1} \mathrm{sr}^{-1} \giga \electronvolt^{-1}]$\\ 
 $3.55 \times 10^3 - 5.01 \times 10^3$ & $4.35 \times 10^3 $  & $(2.94 \pm 0.19)\times 10^{-6}$\\
$5.01 \times 10^3 - 7.08 \times 10^3$ & $6.11 \times 10^3 $  & $(1.13 \pm 0.07)\times 10^{-6}$\\
$7.08 \times 10^3 - 1.00 \times 10^4$ & $8.55 \times 10^3 $  & $(4.73 \pm 0.29)\times 10^{-7}$\\
$1.00 \times 10^4 - 1.41 \times 10^4$ & $1.21 \times 10^4 $  & $(1.94 \pm 0.12)\times 10^{-7}$\\
$1.41 \times 10^4 - 1.99 \times 10^4$ & $1.70 \times 10^4 $  & $(7.95 \pm 0.48)\times 10^{-8}$\\
$1.99 \times 10^4 - 2.82 \times 10^4$ & $2.39 \times 10^4 $  & $(3.19 \pm 0.19)\times 10^{-8}$\\
$2.82 \times 10^4 - 3.98 \times 10^4$ & $3.38 \times 10^4 $  & $(1.28 \pm 0.08)\times 10^{-8}$\\
$3.98 \times 10^4 - 5.62 \times 10^4$ & $4.77 \times 10^4 $  & $(5.07 \pm 0.31)\times 10^{-9}$\\
$5.62 \times 10^4 - 7.94 \times 10^4$ & $6.73 \times 10^4 $  & $(2.05 \pm 0.12)\times 10^{-9}$\\
$7.94 \times 10^4 - 1.12 \times 10^5$ & $9.48 \times 10^4 $  & $(8.29 \pm 0.50)\times 10^{-10}$\\
$1.12 \times 10^5 - 1.58 \times 10^5$ & $1.33 \times 10^5 $  & $(3.40 \pm 0.21)\times 10^{-10}$\\
$1.58 \times 10^5 - 2.23 \times 10^5$ & $1.85 \times 10^5 $  & $(1.43 \pm 0.11)\times 10^{-10}$\\
$2.23 \times 10^5 - 3.16 \times 10^5$ & $2.56 \times 10^5 $  & $(6.24 \pm 0.49)\times 10^{-11}$\\

\end{tabular}
\end{ruledtabular}
\end{table}

\subsection{Systematic uncertainties}

A study of possible systematic effects has been performed. Four main sources of systematic uncertainties on the flux measurement have been considered in this work:  variation of the selection cuts, reliability of the detector simulation, different interaction models, contamination by heavy elements. 

\subsubsection{Selection criteria}
The fiducial selection criteria have been fine tuned in order to obtain an unbiased evaluation of the bayesian probabilities, leading to the best estimation of the cosmic ray proton plus helium energy spectrum. A possible source of systematic error is related to the values of the fiducial cuts on observables used in the event selection procedure. The uncertainty on the measured spectrum has been estimated by applying large variations (about 50 \%) to the fiducial cuts and turns out to be of about 3\%. The bins located at the edges of the measured energy range are affected by an uncertainty of about $\pm 5\%$. A variation of the quality cuts does not give a significative contribution to the total systematic uncertainty. 

\subsubsection{Reliability of the detector simulation}
\begin{figure}[h!]
\includegraphics[width =0.40\textwidth]{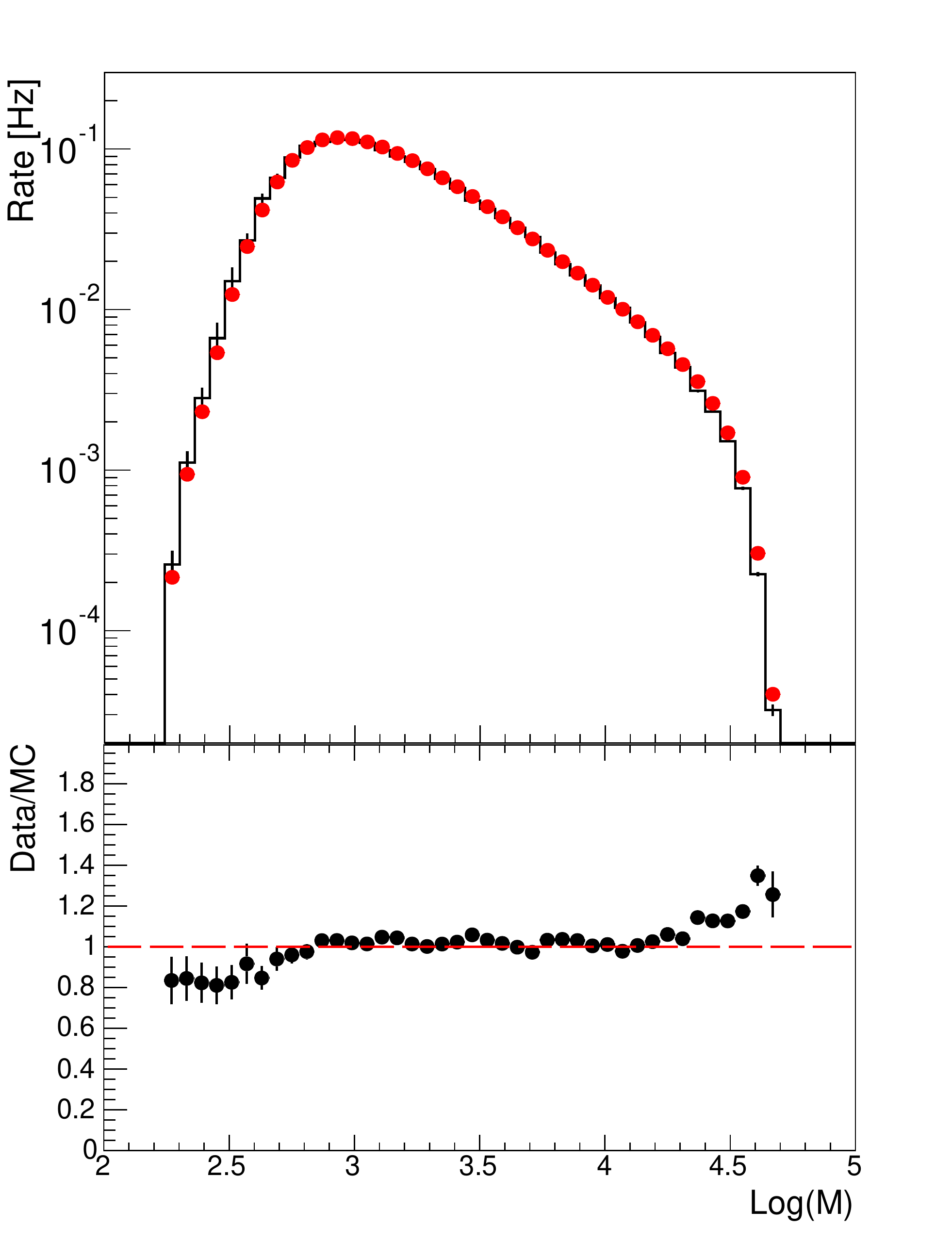}
\caption{Multiplicity distributions of the events selected by the criteria described in section \ref{sect:sel}. Values for data and Monte Carlo (black solid line) are reported. The ratio between data and Monte Carlo is shown in the lower panel.  \label{fig:mult_data_mc}}
\end{figure}

A systematic effect could arise from inaccuracies in the simulation of the detector response. The quality of the simulated events has been estimated by comparing the distribution of the observables obtained by applying the same selection criteria to Monte Carlo simulations and the data sample collected in each different year. As an example in figure \ref{fig:mult_data_mc} the multiplicity distribution obtained from the Monte Carlo events is reported with the multiplicity distribution of the data. The ratio between the two distributions is also reported showing a good agreement between the two distributions. The contribution to the total systematic uncertainty due to the reliability of the detector simulation has been evaluated by using the unfolding probabilities and turns out to be about $\pm 6\%$.

\begin{figure}[h]
\includegraphics[width =0.4\textwidth]{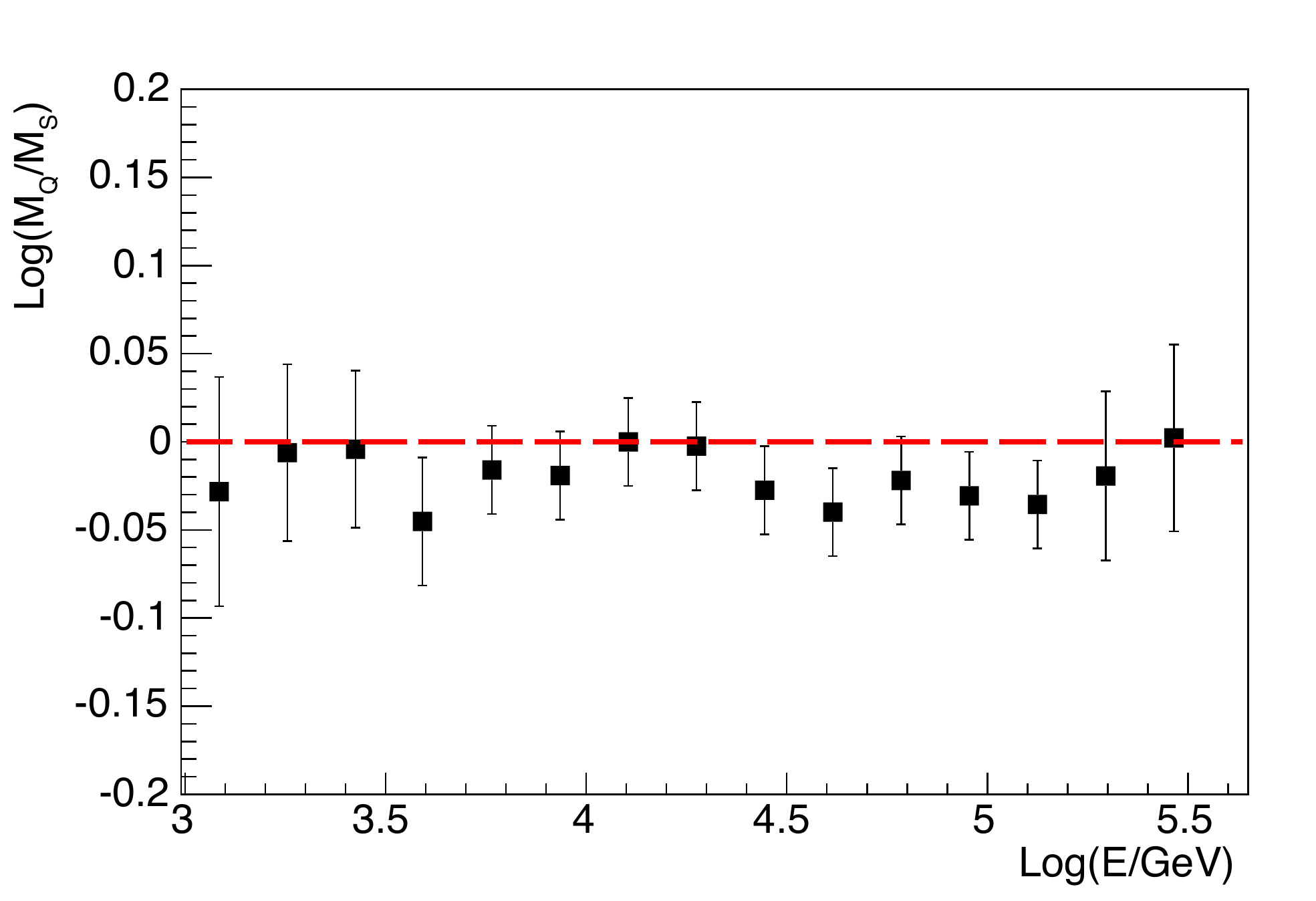}
\caption{Ratio between the multiplicity distributions obtained from QGSJET and SIBYLL based Monte Carlo simulations. \label{fig:qgsjet_syb} }
\end{figure}

\subsubsection{Hadronic interaction models}

In order to estimate effects due to the particular choice of the high energy hadronic interaction model in Monte Carlo simulations, a dataset has been generated by using the SIBYLL 2.1 model \cite{sib1,sib3}. These data have been compared with the QGSJET dataset used in this analysis. In figure \ref{fig:qgsjet_syb} the ratio between the multiplicity distributions obtained by using QGSJET model and the one obtained by using SIBYLL is reported as a function of primary energy. The plot shows that the variation of the multiplicity distributions obtained with the two hadronic models is of few percents, giving a negligible effect on the measured flux.

\subsubsection{Contamination of heavier elements}
\label{sect:cont}
A possible  systematic effect relies in the contamination of elements heavier than Helium. The selection criterion based on the particle density rejects a large fraction of showers produced by heavy primaries, as shown in figure \ref{fig:eff}. The fraction of heavier elements, estimated by using the QGSJET--based simulations according to the H\"orandel model \cite{horandel2003}, is reduced and can be considered as negligible at energies up to $100\, \tera \electronvolt$.  
In the lower energy bins the contamination is about 1\%, whereas in the bins below 100 TeV the contamination does not exceed few \% and in the higher energy bins it is about 10\%. The unfolding procedure has been set up in order to take into account the amount of heavier nuclei passing the selection criteria. The contribution of this effect is therefore not included in the total systematic uncertainty.

\subsubsection{Summary of systematic errors}
The total systematic uncertainty was determined by quadratically adding the individual contributions. The results are affected by a systematic uncertainty of the order of $\pm 5\%$ in the central bins, while the edge bins are affected by a  larger systematic uncertainty less than $\pm 10\%$.

\section{Conclusions}

\label{sect:concl}
The ARGO--YBJ experiment was in operation in its full and stable configuration for more than five years: a huge amount of data has been recorded and reconstructed. The peculiar characteristics of the detector, like the full--coverage technique, high altitude operation and high segmentation and spacetime resolution, allow the detection of showers produced by primaries in a wide energy range from a few $\tera \electronvolt$ up to a few hundreds of $\tera \electronvolt$.  Showers detected by ARGO--YBJ in the multiplicity range $150 - 50000$ strips are mainly produced by primaries in the $(3-300\, \tera \electronvolt)$ energy range. The relation between the shower size spectrum and the cosmic ray energy spectrum has been established by using an  unfolding method based on the Bayes theorem. The unfolding procedure has been performed on the data collected during each year and on the full data sample. The resulting energy spectrum spans the energy range $3-300 \,\tera \electronvolt$, giving a spectral index $\gamma = -2.64 \pm 0.01$, which is in very good agreement with the spectral indices obtained by analyzing the sample collected during each year, therefore demonstrating the excellent stability of the detector over a long period. The resulting spectral indices are also in good agreement with the one obtained by analyzing the first data taken with the detector in its full configuration \cite{argoPRD}. Special care was devoted to the determination of the uncertainties affecting the measured spectrum. The uncertainty on the results is due to systematic effects of the order of $\pm5\%$ in the central energy bins.  This measurement demonstrates the possibility to explore the cosmic ray properties down to the TeV region with a ground--based experiment, giving at present  one of the most accurate measurement of the cosmic ray proton plus helium energy spectrum in the multi--TeV region. 

\begin{acknowledgments}
This work is supported in China by NSFC (Contract No. 101201307940), the Chinese Ministry of Science and 
Technology, the Chinese Academy of Science, the Key Laboratory of Particle Astrophysics, CAS, and in Italy by 
the Istituto Nazionale di Fisica Nucleare (INFN), and Ministero dell'Istruzione, dell'Universit\`a e della
Ricerca (MIUR). We also acknowledge the essential support of W.Y. Chen, G. Yang, X.F. Yuan, C.Y. Zhao, 
R. Assiro, B. Biondo, S. Bricola, F. Budano, A. Corvaglia, B. D'Aquino, R. Esposito, A. Innocente,
A. Mangano, E. Pastori, C. Pinto, E. Reali, F. Taurino and A. Zerbini, in the installation, debugging, and
maintenance of the detector.
\end{acknowledgments}

\bibliography{LightSpectrum_ARGO_PRD_BIB.bib}

\end{document}